\documentclass{ws-ijbc}
\usepackage{ws-rotating}     % used only when sideways tables/figures are used
\usepackage{graphicx}
\usepackage{epstopdf}
\usepackage{multicol}
\begin{document}

\catchline{}{}{}{}{} % Publisher's Area please ignore

\markboth{I.A. Korneev, V.V. Semenov, T.E. Vadivasova}{Synchronization of periodic self-oscillators interacting via memristor-based coupling}

\title{Synchronization of periodic self-oscillators interacting via memristor-based coupling}

\author{Ivan A. Korneev}
\address{Department of Physics, Saratov State University, Astrakhanskaya str., 83\\
Saratov, 410012, Russia\\
ivankorneew@yandex.ru}

\author{Vladimir V. Semenov}

\address{Department of Physics, Saratov State University, Astrakhanskaya str., 83\\
Saratov, 410012, Russia\\
semenov.v.v.ssu@gmail.com}
\address{FEMTO-ST Institute/Optics Department, CNRS \& University Bourgogne Franche-Comt\'e, \\15B avenue des Montboucons
Besan\c con Cedex, 25030, France}

\author{Tatiana E. Vadivasova}
\address{Department of Physics, Saratov State University, Astrakhanskaya str., 83\\
Saratov, 410012, Russia\\
vadivasovate@yandex.ru}

\maketitle

\begin{history}
\received{(to be inserted by publisher)}
\end{history}

\begin{abstract}
A model of two self-sustained oscillators interacting through memristive coupling is studied. The memristive coupling is realized by using a cubic memristor model. Numerical simulation is combined with theoretical analysis by means of quasi-harmonic reduction. It is shown that the specifics of the memristor nonlinearity results in the appearance of infinitely many equilibrium points which form a line of equilibria in the phase space of the system under study. It is established that the possibility to observe the effect of phase locking in the considered system depends on both parameter values and initial conditions. Consequently, the boundaries of the synchronization region are determined by the initial conditions. It is demonstrated that introducing or adding a small term into the memristor state equation gives rise to the disappearance of the line of equilibria and eliminates the dependence of synchronization on the initial conditions.
\end{abstract}


\begin{multicols}{2}

\section{Introduction}
A two-terminal element called "memristor" was initially introduced by Leon Chua as  a realization of a hypothesis of the relationship between the electrical charge and the magnetic flux linkage \cite{chua1971}. Then the idea has been transformed into the conception of ''memristive system'' \cite{chua1976}, which includes the mathematical definition and does not concern the physical sense of dynamical variables and their functional dependence. It allows to combine systems with different nature into one group and to study their properties in a unified manner.  At the present time, the term "memristor" means a two-terminal resistive element, whose resistance (or conductivity) depends on the pre-history of operation. Typically, the current-voltage characteristic of a memristor driven by an external periodic influence represents a pinched hysteresis loop (see for example the characteristic of a cubic memristor model in Fig.\ref{fig1}~(a)). 
In addition to the current-voltage approach, the memristor can be described in the flux-charge domain \cite{corinto2016}. Many experimental prototypes of memristors are known. Development and exploration of such elements are attractive due to their potential applications in electronics and neuroscience \cite{kozma2012,adamatzky2014,tetzlaff2014,radwan2015,vourkas2016,vaidyanathan2017,ventra2013}.  

A memristor attracts attention of specialists in nonlinear dynamics because of its intrinsic properties, which can essentially change the dynamics of electronic oscillatory systems and are responsible for qualitatively new types of the behaviour. There are examples of memristor-based chaotic oscillators \cite{buscarino2012,buscarino2013,pham2013,gambuzza2015-2,zhao2019} and Hamiltonian systems including the memristor \cite{itoh2011,itoh2017}. A variety of effects in memristor oscillators is complemented by the existence of hidden attractors \cite{pham2015,chen2015,chen2015-2} and manifolds of equilibria (in the simplest case it is a line of equilibria) in the phase space \cite{messias2010,botta2011,riaza2012,itoh2008,pham2016,pham2016-2,semenov2015,korneev2017-nd,korneev2017-ch}. 

The issue of the collective dynamics in ensembles of coupled oscillators with memristive coupling is of potential interest from perspective of the nonlinear theory. It represents a distinguished class of problems concerning the influence of adaptive coupling. This topic is attractive in the context of neurodynamics due to an analogy between the memristor dynamics and the behaviour of a neural cell synapse \cite{jo2010,pershin2010-2,williamson2013,li2013,serb2016}. A key step towards understanding of the dynamics of memristively coupled oscillators is to consider the phenomenon of synchronization \cite{pikovsky2001}. There are known publications addressing synchronization of memristively coupled regular \cite{corinto2011,ignatov2016} and chaotic \cite{volos2015,frasca2015,gambuzza2015,zhang2017} self-oscillators. However, results of the mentioned  publications do not allow to reveal distinctive features of this effect as compared to the classical synchronization of self-sustained oscillators coupled via dissipative coupling. In addition, the question on how a particular type of the memristor nonlinearity impacts on the observed effects remains to be actual. Therefore the problem of mutual synchronization of self-sustained oscillators interacting through the memristor is not studied in full. 

In the present work we study synchronization of periodic self-sustained oscillations using as an example two Van der Pol self-sustained oscillators interacting through a memristor. First of all, we aim to answer the question on whether the synchronization through the memristor has intrinsic peculiarities as compared to the synchronization in the case of resistive coupling. We combine our numerical simulation with theoretical analysis by means of quasi-harmonic reduction.

\section{System under study}

\begin{figure*}
\begin{center}
\begin{tabular}{cccc}
\multicolumn{4}{l}{}{\includegraphics[width=2.0\columnwidth]{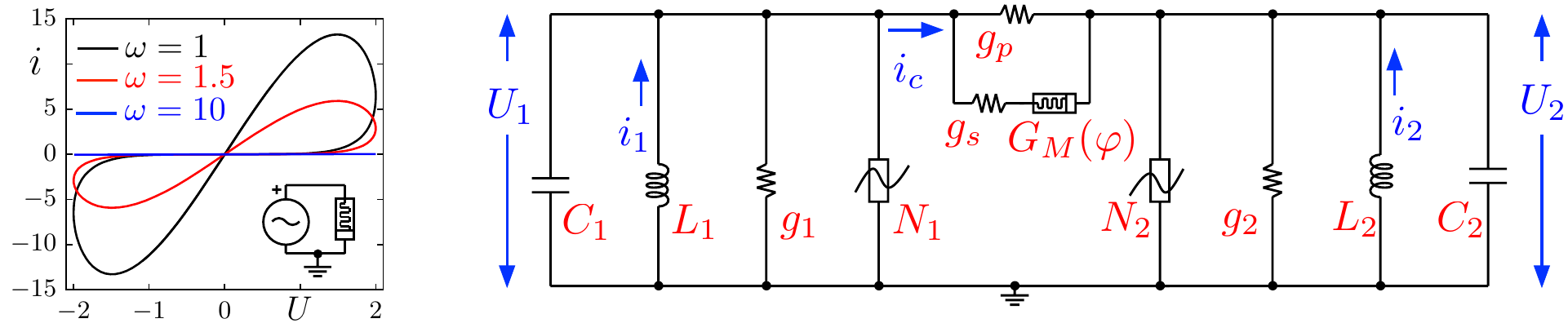}}\\
\multicolumn{4}{l}{}{\hspace{0.0cm} (a) \hspace{8.0cm} (b) \hspace{5.5cm}} \\
\end{tabular}
\end{center}
\caption{(a) Current-voltage characteristic of the cubic memristor model (\ref{cubic_memristor}) with the parameters $\mu=0.02$, $\nu=0.8$ driven by the periodic voltage signal $U_{ext}=2 \sin{(\omega t)}$ at $\omega=1$ (black line), $\omega=1.5$ (red line), $\omega=10$ (blue line). (b) Schematic circuit diagram of the system under study (Eqs.(\ref{physical})).}
\label{fig1}                                                                                                   
\end{figure*}

According to the paper \cite{chua1971} the memristor relates the transferred electrical charge, $q(t)$, and the magnetic flux linkage, $\varphi(t)$: $dq=G_{M}d\varphi$. The following dependence is assumed to be satisfied in a model of the cubic memristor: $q(\varphi)=\mu\varphi+\frac{1}{3}\nu\varphi^{3}$. Then we have
\begin{equation}
G_{M}=G_{M}(\varphi)=\dfrac{dq}{d\varphi}=\mu+ \nu\varphi^{2}.
\label{cubic_memristor}
\end{equation}
In the following, the variable $\varphi$ is considered as a state variable defined mathematically as $\varphi(t)~=~\int\limits_{-\infty}^{t}{U(t)dt}$ and is not associated with the magnetic field. By using the formulas $d\varphi=Udt$ and $dq=idt$ ($U$ is the voltage across the memristor, $i$ is the current passing through the memristor) the memristor current-voltage equation can be derived: $i=G_{M}(\varphi)U$. It means that $G_{M}$ is the conductance (memductance) and depends on the entire past history of $U(t)$: $G_{M}(\varphi)=G_{M}\left(\int\limits_{-\infty}^{t}{U(t)dt}\right)$. The model of the cubic memristor forced by the periodic voltage signal $U_{ext}=A_{ext} \sin{(\omega t)}$ exhibits the current-voltage characteristic, $i(U)$, which depicts a hysteresis loop being pinched in the case of increasing frequency $\omega$ [Fig.\ref{fig1}~(a)]. 

The system under study is pictured in Fig. \ref{fig1}(b). It consists of two coupled self-sustained oscillators. Each partial self-oscillator represents a parallel oscillatory circuit including the capacitor $C$, the inductor $L$, a resistor with the conductance $g$, and the nonlinear element $N$ with the N-type current-voltage characteristic described by the formula: $i(U) = -\alpha U + \beta U^{3}$. The dynamics of the partial self-sustained oscillator is described by the Van der Pol self-sustained oscillator. The memristive coupling is realized by the cubic memristor (\ref{cubic_memristor}) with the conductance $G_{M}(\varphi)$ and additional resistors with conductances $g_{p}$ and $g_{s}$. Adjusting the conductances $g_{p}$ and $g_{s}$, one can change the summary conductance, which is responsible for the coupling strength and can be presented in the form $kG_{M}(\varphi)$. By using the Kirchhoff's current law the following differential equations for the voltages $U_{1,2}$ across the capacitors $C_{1,2}$ and the currents $i_{1,2}$ through the inductances $L_{1,2}$ can be derived:

\begin{equation}
\label{physical}
\left\lbrace
\begin{array}{l}
\dfrac{d U_{1}}{dt_{*}} + \dfrac{1}{C_{1}}i_1 + \dfrac{g_1}{C_1}U_1 + \dfrac{kG_M(\varphi)}{C_1}(U_1 - U_2)\\
- \dfrac{\alpha_1}{C_1}U_1 + \dfrac{\beta_1}{C_1}U_1^3 = 0, \\
\\
\dfrac{d U_2}{dt_{*}} + \dfrac{1}{C_2}i_2 + \dfrac{g_2}{C_2}U_2 +  \dfrac{kG_M(\varphi)}{C_2}(U_2 - U_1)\\
- \dfrac{\alpha_2}{C_2}U_2 + \dfrac{\beta_2}{C_2}U_2^3 = 0, \\
\\
\dfrac{1}{C_{1}}\dfrac{di_{1}}{dt_{*}}=\dfrac{1}{C_{1}L_{1}}U_{1}, \\
\\
\dfrac{1}{C_{2}}\dfrac{di_{2}}{dt_{*}}=\dfrac{1}{C_{2}L_{2}}U_{2}, \\
\\
\dfrac{d\varphi}{dt_{*}}= U_1 - U_2, 
\end{array}
\right.
\end{equation}
where $t_{*}$  is the physical time.
The following parameters are assumed to be equal: $\alpha_{1}=\alpha_{2}=\alpha$, $\beta_{1}=\beta_{2}=\beta$, $g_{1}=g_{2}=g$, $C_{1}=C_{2}=C$. Let us denote $\omega_1^2 = \frac{1}{L_1 C}$ and $\omega_2^2 = \frac{1}{L_2 C}$, $p=\omega_{1}^{2}/ \omega_{2}^{2}$ and introduce the dimensionless time and variables as follows:  

\begin{align}\label{normalization}
& t= \omega_{1}t_{*}, \quad x_{1} = \sqrt{\dfrac{\beta}{C\omega_1}}U_1, \quad x_{2}=\sqrt{\dfrac{\beta}{C\omega_1}}U_2, \nonumber \\
& y_{1} = \dfrac{1}{\omega_{1}C}\sqrt{\dfrac{\beta}{C\omega_1}}i_{1}, \quad y_{2} = \dfrac{1}{p\omega_{1}C}\sqrt{\dfrac{\beta}{C\omega_1}}i_{2}, \nonumber \\
& z = \omega_{1}\sqrt{\dfrac{\beta}{C\omega_1}}\varphi.
\end{align}
Then the system (\ref{physical}) reads
\begin{equation}
\label{system}
\left\lbrace
\begin{array}{l}
\dot{x}_1 + y_1 - \left(\gamma - x_{1}^{2}\right)x_1 + k G(z)(x_1 - x_2) = 0,\\
\dot{x}_2 + p y_2 - \left(\gamma - x_2^2\right)x_2 + k G(z)(x_2 - x_1) = 0,\\
\dot{y}_1 = x_1,\\
\dot{y}_2 = x_2,\\
\dot{z} = x_1 - x_2,
\end{array}
\right.
\end{equation}
where $\dot{x}_{1,2}=\frac{dx_{1,2}}{dt}$, $\dot{y}_{1,2}=\frac{dy_{1,2}}{dt}$, $\gamma = \frac{\alpha-g}{C\omega_{1}}$, $G(z)~=~\frac{\mu+ \nu\varphi^{2}}{C\omega_1}~=~a+bz^2$. The equilibrium points of the system (\ref{system}) have coordinates $x_{1,2}=0$, $y_{1,2}=0$, $z\in(-\infty;\infty)$. It means that the system (\ref{system}) has a line of equilibria in its phase space, i.e., each point on the axis OZ is an equilibrium point.

The dynamical variable $z$ can be excluded from the system (\ref{system}). Indeed, it results from the last equation that $\dot{z} = \dot{y}_1 - \dot{y}_2$. Then one can derive $z(t) = z(0)+y_{1}(t) - y_{2}(t) - y_{1}(0) + y_{2}(0)$. This implies that the value of the memristor conductance at any time depends on both the instantaneous values $y_{1}$ and $y_{2}$ and the initial values $y_{1}(0)$, $y_{2}(0)$ and $z(0)$. Hence, it follows that the system (\ref{system}) describes two interacting self-oscillators with dissipative coupling, whose strength depends on both the instantaneous and initial values of the dynamical variables:
\begin{equation}
\label{system2}
\left\lbrace
\begin{array}{l}
\dot{x}_1 + y_1 - \left(\gamma - x_{1}^{2}\right)x_1 \\
= k G(z(0)+y_1 - y_2 - y_{1}(0) + y_{2}(0))(x_2 - x_1),\\
\dot{x}_2 + p y_2 - \left(\gamma - x_2^2\right)x_2 \\
= k G(z(0)+y_1 - y_2 - y_{1}(0) + y_{2}(0))(x_1 - x_2),\\
\dot{y}_1 = x_1,\\
\dot{y}_2 = x_2.\\
\end{array}
\right.
\end{equation}
It gives rise to the possibility to control the coupling strength by changing the initial conditions. By this way one can induce (or destroy) the effect of phase locking at fixed values of the parameters. 
\section{Results}
The system of two coupled Van der Pol self-oscillators (\ref{system}) is considered at fixed parameters of the memristor characteristic $a=0.02$, $b=0.8$ and the self-oscillation excitation parameter $\gamma=0.1$. The coupling strength, $k$, and the frequency mismatch parameter, $p$, are varied.
\subsection{Numerical modelling}
Numerical simulations were carried out by integrating Eqs. (\ref{system}) using the Runge-Kutta fourth-order method with the time step $\Delta t = 0.001$. The numerically obtained time realizations were used for plotting phase portraits and calculating the instantaneous phase of self-oscillations in the partial systems. The instantaneous phases of self-oscillators $\Psi_{1}(t)$ and $\Psi_{2}(t)$ are defined as follows

\begin{equation}
\label{phase}
\Psi_{i}(t)=\text{arctg} {\frac{y_{i}(t)}{x_{i}(t)}} \pm \pi n(t),~~ i=1,2,
\end{equation}
where $n(t)$ is an integer variable defined by the condition of phase continuity. Using the instantaneous phases, one can determine the phase difference $\Delta \Psi(t)= \Psi_{2}(t)-\Psi_{1}(t)$ and the mean difference frequency:
\begin{equation}
\label{Omega1}
\Omega=\lim_{T\rightarrow \infty}{\frac{\Delta \Psi(t+T)-\Delta \Psi(t)}{T}}.
\end{equation}
It is evident that the quantity $\Omega$ vanishes in the synchronization region. In the case of large coupling strength, the synchronization can be realized through suppression of self-oscillations of either self-oscillator. Then calculating $\Omega$ through the formula (\ref{Omega1}) gives rise to incorrect results.

The following results have been obtained by numerical modelling the system (\ref{system}) without the frequency mismatch ($p=1$).
The in-phase regime of synchronization corresponding to $x_{1}(t) \equiv x_{2}(t)$, $y_{1}(t) \equiv y_{2}(t)$ is achieved at any positive values of the coupling strength $k>0$ and for any initial conditions. In this case the interaction through a memristor leads to the same phenomenon as compared to usual dissipative coupling. The difference takes place only in the context of transient time. The transient process duration in the system (\ref{system}) strongly depends on initial conditions. 

\begin{figure*}
\begin{center}
\begin{tabular}{cc}
\includegraphics[width=0.49\columnwidth]{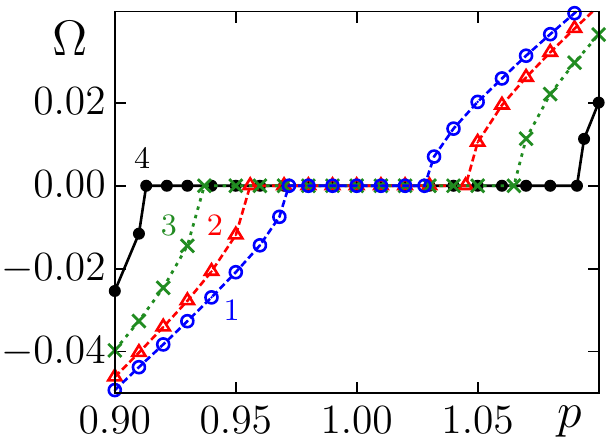} & \includegraphics[width=0.474\columnwidth]{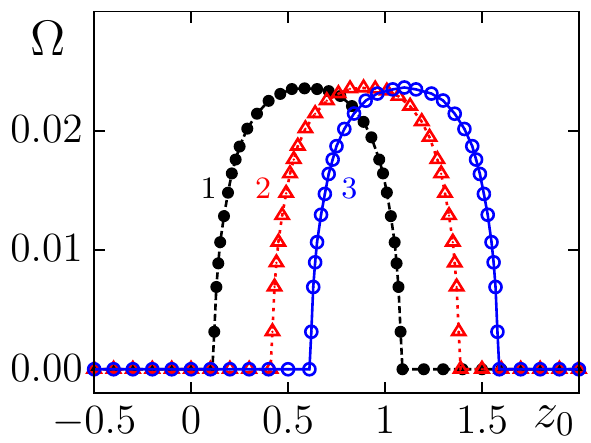} \\
(a) & (b) \\
\end{tabular}
\end{center}
\caption{System (\ref{system}) in numerical experiments. (a) Dependence of the mean difference frequency $\Omega$ on the frequency mismatch $p$ for different initial values $z_{0}$: $z_{0}=0$ (blue curve 1), $z_{0}=-0.25$ (red curve 2), $z_{0}=-0.5$ (green curve 3), $z_{0}=-0.75$ (black curve 4). Parameters are: $\gamma=0.1$, $k=0.02$, $a=0.02$, $b=0.8$. Other initial conditions are $x_{1}(0)=0.5$, $y_{1}(0)=0.5$, $x_{2}(0)=-0.5$, $y_{2}(0)=-0.4$. (b) Dependence of the mean difference frequency $\Omega$ on the initial value $z(0)=z_0$ for different sets of initial values of the other dynamical variables: $x_{1}(0)=0.5$, $y_{1}(0)=0.5$, $x_{2}(0)=-0.3$, $y_{2}(0)=-0.1$ (black curve 1), $x_{1}(0)=0.5$, $y_{1}(0)=0.5$, $x_{2}(0)=-0.5$, $y_{2}(0)=-0.4$ (red curve 2), $x_{1}(0)=0.5$, $y_{1}(0)=0.5$, $x_{2}(0)=-0.2$, $y_{2}(0)=-0.6$ (blue curve 3). Parameters are: $\gamma=0.1$, $p=1.05$, $k=0.1$, $a=0.02$, $b=0.8$.}
\label{fig2}                                                                                                   
\end{figure*}

\begin{figure*}[t]
\begin{center}
\begin{tabular}{cc}
\includegraphics[width=0.47\columnwidth]{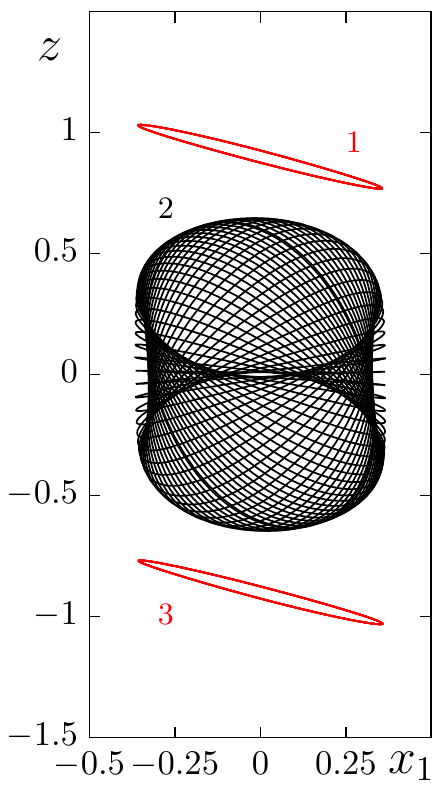} & \includegraphics[width=0.50\columnwidth]{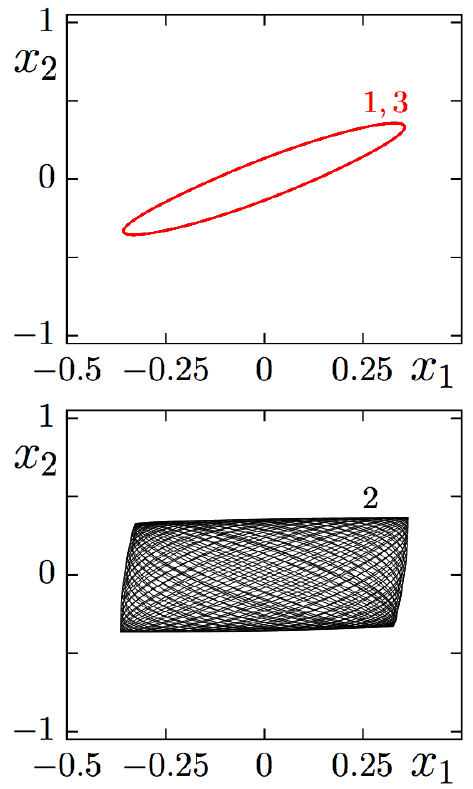} \\
(a) & (b) \\
\end{tabular}
\end{center}
\caption{Phase trajectories of the system (\ref{system}) in the ($x_{1}$,$z$) (the panel (a)) and ($x_{1}$,$x_{2}$) (the panel (b)) planes. The red curves correspond to the regime of synchronization, the black trajectory traces quasi-periodical oscillations. Initial conditions are: $x_{1}(0)=0.5$, $y_{1}(0)=0.5$, $x_{2}(0)=-0.3$, $y_{2}(0)=-0.1$, $z(0)=1.5$ (curve 1), $x_{1}(0)=0.5$, $y_{1}(0)=0.5$, $x_{2}(0)=-0.3$, $y_{2}(0)=0.1$, $z(0)=0.4$ (curve 2), $x_{1}(0)=0.5$, $y_{1}(0)=0.5$, $x_{2}(0)=-0.5$, $y_{2}(0)=-0.4$, $z(0)=0.0$ (curve 3). Parameters are: $p=1.05$, $k=0.1$,  $\gamma=0.1$, $a=0.02$, $b=0.8$.}
\label{fig3}                                                                                                   
\end{figure*}
Let us consider the system (\ref{system}) in the presence of weak frequency mismatch ($p \neq 1$). The instantaneous phases of the partial self-oscillators and the mean difference frequency $\Omega$ (see the formula (\ref{Omega1})) were calculated to detect mutual phase and frequency locking. The dependence $\Omega(p)$ enables one to reveal the phase and frequency locking effect and to estimate the synchronization region. It has been shown numerically that the memristive coupling provides an opportunity for mutual locking of the phases and frequencies of self-oscillators similarly to usual dissipative coupling. There is a certain interval of the frequency mismatch where the mean difference frequency $\Omega$ equals to zero. This effect was observed in electronic experiments described in the paper \cite{ignatov2016}. However, the synchronization via the memristor has an essential feature: the width of the phase-frequency locking region continuously depends on the initial conditions. Numerically obtained dependences $\Omega (p)$ corresponding to the fixed coupling strength $k=0.02$ and different initial values of the variable $z(0)=z_{0}$ are depicted in Fig.\ref{fig2} (a). It is seen that the synchronization region boundaries are essentially different for different values $z_{0}$. The width of the synchronization area increases with the growth of the absolute value $|z_{0}|$. The dependence of the mean difference frequency on the initial value $z_{0}=z(0)$ indicates the influence of initial conditions [Fig. \ref{fig2} (b)]. It was calculated for the fixed parameters $p=1.05$, $k=0.1$ and different initial values of the other variables. On each curve $\Omega (z_{0})$ depicted in Fig. \ref{fig2} (b) one can distinguish an interval of values $z_{0}$, where the effect of synchronization is not observed. The boundaries of this interval are varied or change depending on the initial values of the other dynamical variables, but all the curves presented in Fig. \ref{fig2} (b) have an identical shape.

Projections of phase trajectories corresponding to the existence or absence of synchronization are shown in Fig. \ref{fig3}. The trajectories were obtained from different initial conditions at the same parameter values. Two identical red closed curves in Fig. \ref{fig3} illustrate synchronous oscillations. Projections of the synchronous oscillations are identical in the space of variables $x_1$, $x_2$, $y_1$, $y_2$ (see for example Fig. \ref{fig3} (b)). However, there is a shift along the $OZ$ axis in the full phase space (compare curves 1 and 3 in Fig. \ref{fig3} (a)). Projections of non-synchronous oscillations (the black trajectory in Fig.~\ref{fig3}) trace a figure being topologically equivalent to a two-dimensional torus. The figures obtained from different initial conditions have a different shape. 

The results presented above have shown that the possibility to observe the regime of synchronization in the system (\ref{system}) depends on the initial conditions. After that the question can be arised: Whether characteristics of synchronous and non-synchronous oscillations continuously depend on initial conditions? It is known that a continuous dependence of oscillation characteristics on initial conditions is typical for oscillators with a line of equilibria including memristor-based oscillators \cite{messias2010,botta2011,semenov2015,korneev2017-ch, korneev2017-nd}. Therefore one can assume that the synchronization region boundaries continuously depend on the initial conditions in some area on the ($p,k$) plane. However, this assumption requires a detailed theoretical analysis of the model (\ref{system}).

\subsection{Theoretical analysis}
Self-oscillations in the partial self-oscillators of the system (\ref{system}) are close to harmonic at small positive values of the parameter $\gamma$. In such a case one can derive reduced equations for the instantaneous amplitude and phase by applying the Van der Pol method. In terms of quasi-harmonic reduction the solution of Eqs. (\ref{system}) is found in the following form:

\begin{eqnarray}
\label{zamena}
y_{1,2}(t)&=&\text{Re}\Big[ a_{1,2}(t)e^{jt}\Big]=\frac{1}{2} \left( a_{1,2}(t)e^{jt}+a_{1,2}^{*}(t)e^{-jt} \right), \nonumber \\
x_{1,2}(t)&=&\frac{j}{2} \left( a_{1,2}(t)e^{jt}-a_{1,2}^{*}(t)e^{-jt} \right), 
\end{eqnarray}
where $a_{1}(t)$ and $a_{2}(t)$ are the instantaneous complex amplitudes of self-oscillations in the partial self-oscillators, $a^{*}_{1}(t)$ and $a^{*}_{2}(t)$ are the complex conjugate functions, $j$ is the imaginary unit. The amplitudes $a_{1}(t)$ and $a_{2}(t)$ are assumed to be slowly varying during the period of self-oscillations $T_{0}=2\pi$. In addition, the following condition for the first derivatives is assumed to be satisfied: $\dot{a}_{1,2}e^{jt}+\dot{a}_{1,2}^{*}e^{-jt}=0$.  The equation for the variable $z(t)$ can be derived by using the last equation of the system (\ref{system}) and the substitution (\ref{zamena}):

\begin{eqnarray}
\label{z}
z(t)&=&z(0)+\int_{0}^{t}{(x_{1}(\tau)-x_{2}(\tau))d \tau}\nonumber \\
&=& z(0)+y_{1}(t)-y_{2}(t)+y_{2}(0)-y_{1}(0) \nonumber \\
&=& C_{0}+\frac{1}{2}(a_{1}-a_{2})e^{jt}+\frac{1}{2}(a_{1}^{*}-a_{2}^{*})e^{-jt},
\end{eqnarray}
where $C_{0} = z(0)+y_{2}(0)-y_{1}(0)$ is a constant determined by the initial state of the system. Next, the expressions (\ref{zamena}) and (\ref{z}) are 
inserted into Eqs. (\ref{system}). Then a system of equations for the complex amplitudes is derived by using the memristor characteristic and the condition for the derivatives. After averaging of the complex amplitudes and their derivatives over the period $T_0$ the following system of reduced equations is developed:
\begin{equation}
\label{a1-a2}
\begin{array}{l}
\dot{a}_{1} = \frac{\gamma}{2}a_{1} - \frac{3}{8}a_{1}|a_{1}|^{2}+
\frac{k}{2}(a+bC_{0}^{2})(a_{2}-a_{1})\\
\\
+\frac{kb}{8}|a_{2}-a_{1}|^{2}(a_{2}-a_{1}), \\
\\
\dot{a}_{2} = \frac{\gamma}{2}a_{2} - \frac{3}{8}a_{2}|a_{2}|^{2}+ \frac{j(p-1)}{2}a_{2}\\
\\
+\frac{k}{2}(a+bC_{0}^{2})(a_{1}-a_{2})+\frac{kb}{8}|a_{1}-a_{2}|^{2}(a_{1}-a_{2}).
\end{array}
\end{equation}

The system (\ref{a1-a2}) is presented as a system of equations for real amplitudes $A_{1}$, $A_{2}$ and phases 
$\phi_{1}$, $\phi_{2}$ \footnote{More precisely, the variable $\phi_{i}$ is a slow-varying component of the full phase of self-oscillations $\Phi_{i}=t+\phi_{i}$, $i=1,2$.} by using the substitution $a_{1,2}=A_{1,2}\exp{[j \phi_{1,2}]}$:
\begin{equation}
\label{A-phi}
\begin{array}{l}
\dot{A}_{1} = \frac{\gamma}{2}A_{1}-\frac{3}{8}A_{1}^{3}+\frac{k}{2} \Big[ a+b\Big( C_{0}^{2}+ \frac{A_{1}^{2}+A_{2}^{2}}{4}\\
-\frac{A_{1}A_{2}}{2}\cos{(\phi_{2}-\phi_{1})} \Big) \Big]
(A_{2}\cos{(\phi_{2}-\phi_{1})}-A_{1}), \\
\\
\dot{\phi}_{1} = \frac{k}{2} \Big[ a+b\Big( C_{0}^{2}+ \frac{A_{1}^{2}+A_{2}^{2}}{4}\\
-\frac{A_{1}A_{2}}{2}\cos{(\phi_{2}-\phi_{1})} \Big) \Big] 
\frac{A_{2}}{A_{1}}\sin{(\phi_{2}-\phi_{1})}, \\
\\
\dot{A}_{2} = \frac{\gamma}{2}A_{2}-\frac{3}{8}A_{2}^{3}+\frac{k}{2}
\Big[ a+b\Big( C_{0}^{2}+ \frac{A_{1}^{2}+A_{2}^{2}}{4}\\
-\frac{A_{1}A_{2}}{2}\cos{(\phi_{2}-\phi_{1})} \Big) \Big] 
(A_{1}\cos{(\phi_{2}-\phi_{1})}-A_{2}), \\
\\
\dot{\phi}_{2} = \frac{p-1}{2}-\frac{k}{2} \Big[ a+b\Big( C_{0}^{2}+ \frac{A_{1}^{2}+A_{2}^{2}}{4}\\
-\frac{A_{1}A_{2}}{2}\cos{(\phi_{2}-\phi_{1})} \Big) \Big]  
\frac{A_{1}}{A_{2}}\sin{(\phi_{2}-\phi_{1})}.
\end{array}
\end{equation}
Introducing of the phase difference $\theta=\phi_{2}-\phi_{1}$ allows to rewrite Eqs. (\ref{A-phi}) as follows: 
\begin{equation}
\begin{array}{l}
\label{A-theta}
\dot{A}_{1} = \frac{\gamma}{2}A_{1}-\frac{3}{8}A_{1}^{3}+\frac{k}{2}
\Big[ a+b\Big( C_{0}^{2}+ \frac{A_{1}^{2}+A_{2}^{2}}{4}\\
-\frac{A_{1}A_{2}}{2}\cos{\theta} \Big) \Big](A_{2}\cos{\theta}-A_{1}), \\
\\
\dot{A}_{2} = \frac{\gamma}{2}A_{2}-\frac{3}{8}A_{2}^{3}+\frac{k}{2}
\Big[ a+b\Big( C_{0}^{2}+ \frac{A_{1}^{2}+A_{2}^{2}}{4}\\
-\frac{A_{1}A_{2}}{2}\cos{\theta} \Big) \Big](A_{1}\cos{\theta}-A_{2}), \\
\\
\dot{\theta} = \Delta-\frac{k}{2}\Big[ a+b\Big( C_{0}^{2}+ \frac{A_{1}^{2}+A_{2}^{2}}{4}\\
-\frac{A_{1}A_{2}}{2}\cos{\theta} \Big) \Big] \left( \frac{A_{1}}{A_{2}}+\frac{A_{2}}{A_{1}} \right)\sin{\theta},
\end{array}
\end{equation}
where $\Delta = \frac{p-1}{2}$.

Next, the phase reduction is used to describe coupled self-oscillators. It means that the real amplitudes of the self-oscillators are assumed to be almost constant and equal to a stationary value in the absence of coupling: 
\begin{eqnarray}
\label{A0}
A_{1}=A_{2}=A_{0}=\sqrt{\frac{4\gamma}{3}}. 
\end{eqnarray}
Using (\ref{A0}) and (\ref{A-theta}) we obtain the equation for the phase difference: 
\begin{equation}
\begin{array}{l}
\label{theta}
\dot{\theta} = \Delta - kF(\theta) = \Delta \\
- k\left[ a+b\left( C_{0}^{2}+ \frac{A_{0}^{2}}{2}-\frac{A_{0}^{2}}{2}\cos{\theta} \right) \right]\sin{\theta}.
\end{array}
\end{equation}
If the expression in the square brackets changes to a constant  $\eta$, then Eq. (\ref{theta}) is transformed to the Adler equation describing synchronization of quasi-harmonic self-oscillators with dissipative coupling:
\begin{eqnarray}
\label{Adler}
\dot{\theta} = \Delta - \Delta_{s}\sin{\theta},~~~ \Delta_{s}=k \eta.
\end{eqnarray}
The phase synchronization area corresponds to the existence of the stable solution $\theta_{0}=const$. In this case the difference frequency becomes $\Omega = \dot{\theta} \equiv 0$. Then the boundaries of the synchronization area can be found in the case of Eq.~(\ref{Adler}):
\begin{eqnarray}
\label{boundary}
|\Delta| \leq \Delta_{s}.
\end{eqnarray}
Outside the synchronization region the mean difference frequency is determined by the known formula:
\begin{eqnarray}
\label{Omega}
\Omega = <\dot{\theta}> = \sqrt{\Delta^{2}-\Delta_{s}^{2}}, ~~~|\Delta| \geq \Delta_{s}.
\end{eqnarray}
Here the brackets $<...>$ mean the time-averaging operation. In the case of Eq. (\ref{theta}) the boundaries of the synchronization area cannot be calculated analytically. However, the term of Eq. (\ref{theta}) including $\cos{\theta}$ can be neglected in the case of a large absolute value of the constant $C_{0}$. Then Eqs. (\ref{boundary}) and (\ref{Omega}) are true and include the parameter $\Delta_{s}$ determined by the formula:
\begin{eqnarray}
\label{Dels}
\Delta_{s} \approx k\left[ a+b\left( C_{0}^{2}+ \frac{2\gamma}{3} \right) \right].
\end{eqnarray}

\begin{figure*}[t]
\begin{center}
\begin{tabular}{cc}
\includegraphics[width=0.5\columnwidth]{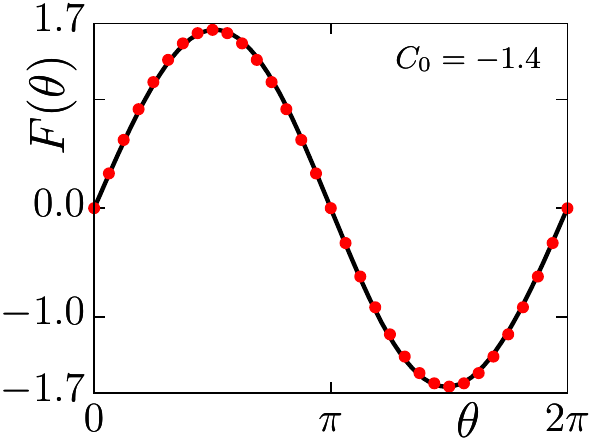} & \includegraphics[width=0.5\columnwidth]{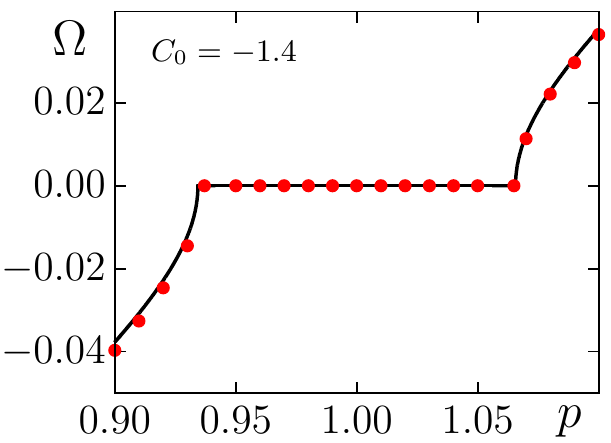} \\
\includegraphics[width=0.5\columnwidth]{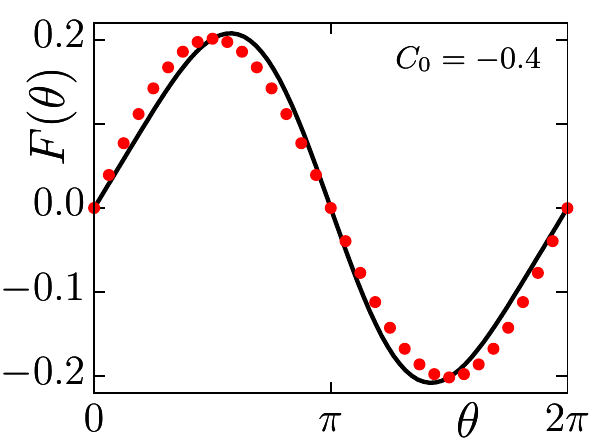} & \includegraphics[width=0.5\columnwidth]{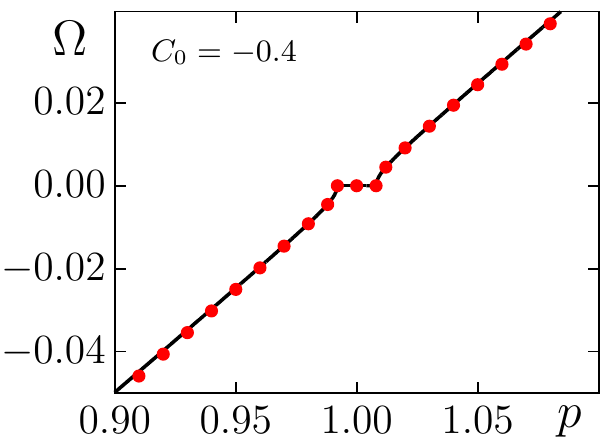} \\
(a) & (b) \\
\end{tabular}
\end{center}
\caption{(a) Comparison of the function $F(\theta)$ of Eq. (\ref{theta}) (red  circles) at $C_{0}=-1.4$ (the upper panel) and $C_{0}=-0.4$ (the lower panel) and the function $\eta\sin(\theta)$ (black line); (b) Dependences of the mean difference frequency on the frequency mismatch $\Omega(p)$. Numerical results are shown by red circles and the corresponding theoretical curves (see Eq. (\ref{Omega})) are coloured in black. The initial conditions for the $z$ variable are $z(0)=-0.5$ (the upper panel) and $z(0)=0.5$ (the lower panel). Other initial conditions are $x_{1}(0)=0.5$, $y_{1}(0)=0.5$, $x_{2}(0)=-0.5$, $y_{2}(0)=- 0.4$. Parameters of the system (\ref{system}) are $\gamma=0.1$, $a=0.02$, $b=0.8$, $k=0.02$.}
\label{fig4}                                                                                                   
\end{figure*}

Figure \ref{fig4} (a) illustrates functions $\eta \sin{(\theta)}$ and $F(\theta)$, which defines the right part of Eq. (\ref{theta}). For the chosen initial conditions corresponding to $C_{0}=-1.4$ [Fig. \ref{fig4} (a), the upper panel] and $C_{0}=-0.4$ [Fig. \ref{fig4} (a), the lower panel] the curves are almost identical. It allows to use the condition (\ref{boundary}) and the formula (\ref{Omega}) for estimating the dependence $\Omega(p)$. Theoretical results and numerical findings for the system (\ref{system}) are presented in Fig. \ref{fig4} (b). For the chosen set of the parameters and  initial conditions the similarity between the results of numerical modelling and analytical approach is evident. 

\begin{figure*}[t]
\begin{center}
\begin{tabular}{cc}
\includegraphics[width=0.48\columnwidth]{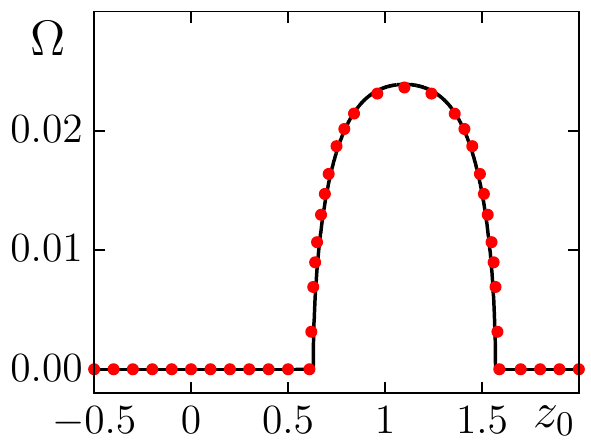} & \includegraphics[width=0.48\columnwidth]{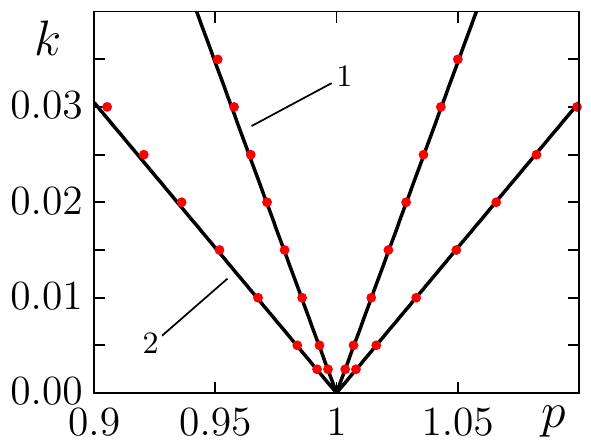} \\
(a) & (b) \\
\end{tabular}
\end{center}
\caption{Numerical and theoretical results for the system (\ref{system}). (a) Dependence of the mean difference frequency $\Omega$ on the initial value $z(0)$ for the parameters $p=1.05$, $k=0.1$, $\gamma=0.1$ and other initial conditions $x_{1}(0)=0.5$, $y_{1}(0)=0.5$, $x_{2}(0)=-0.2$, $y_{2}(0)=-0.6$. Numerical results are shown by red circles and the corresponding theoretical curve (see Eq. (\ref{Omega})) is coloured in black. (b) Boundaries of the synchronization region obtained numerically (red circles) and theoretically (black solid lines). The theoretical curves are calculated according to the formula (\ref{boundary}). Boundaries 1 correspond to the initial conditions $x_1=0.5$, $y_1(0)=0.5$, $x_2(0)=-0.5$, $y_2(0)=-0.4$, $z(0)=0$. Boundaries 2 are constructed for the initial conditions $x_1(0)=0.5$, $y_1(0)=0.5$, $x_2(0)=-0.5$, $y_2(0)=-0.4$, $z(0)=-0.5$. The self-oscillation excitation parameter is $\gamma=0.1$.}
\label{fig5}                                                                                                   
\end{figure*}

Figure \ref{fig5} also demonstrates a good correspondence between results of numerical experiments the numerical results and the theoretical data. Figure \ref{fig5} (a) shows the dependence of the mean difference frequency $\Omega$ on the initial value $z(0)=z_{0}$ obtained numerically and analytically for the system (\ref{system}). The boundaries of the synchronization region obtained by using the condition (\ref{boundary}) for two values of the constant $C_{0}$ (the solid lines in Fig. \ref{fig5} (b)) are close to the numerically estimated ones for the same initial conditions (the red circles in Fig. \ref{fig5} (b)). If the value of the constant $C_{0}$ is close to zero, then the term in Eq. (\ref{theta}) including $\cos(\theta)$ cannot be neglected. As a result, visible difference appears between the results of numerical modelling and theoretical approach involving the formulas (\ref{boundary}) and (\ref{Omega}). 
Nevertheless, the theoretical results presented above allow to conclude that a continuous variation of the initial conditions $y_{1}(0)$, $y_{2}(0)$  and $z(0)$ in certain intervals gives rise to a continuous change in the quantity $\Delta_{s}$ (see Eq. (\ref{Dels})) and in the boundaries of the phase locking area. The numerical findings confirm this fact. 

\section{Role of the memristor state equation}
The appearance of a line of equilibria in the phase space of the system (\ref{system}) results from peculiarities of the memristor state equation (the last equation of the system (\ref{system})). The existence of the line of equilibria is a non-robust effect. It is difficult to imagine its implementation in real physical systems, which inevitably include sources of fluctuations and have their own intrinsic peculiarities \cite{semenov2015,korneev2017-ch, korneev2017-nd}. 
The state equation of the system (\ref{system}) is one of the simplest forms and follows from the initial Chua's introduction of the memristor \cite{chua1971}. In general, the memristor state equation can be more complex \cite{chua1976}. Let us consider how the change in the memristive coupling element model affects the studied phenomenon. Further consideration of the system (\ref{system}) is carried out for the modified last equation in the following form:
\begin{eqnarray}
\label{equation-z2}
\dot z = x_1 - x_2 - \delta z,
\end{eqnarray}
where $\delta$ is a small parameter. Change of configuration of the memristor state equation results in disappearance of the line of equilibria at any non-zero value of the parameter $\delta$. There is one point of equilibrium in the phase space of the system with the modified last equation. Stability of the equilibrium point is determined by a sign of the parameter $\delta$. In case $\delta > 0$ perturbations along the axis $OZ$ are damped and stationary regimes do not depend on initial conditions. In case $\delta < 0$ the perturbations along the axis $OZ$ increase during time of observation and trajectories tend to $\pm \infty$ along the axis $OZ$.

\begin{figure*}[t]
\begin{center}
\begin{tabular}{cc}
\includegraphics[width=0.49\columnwidth]{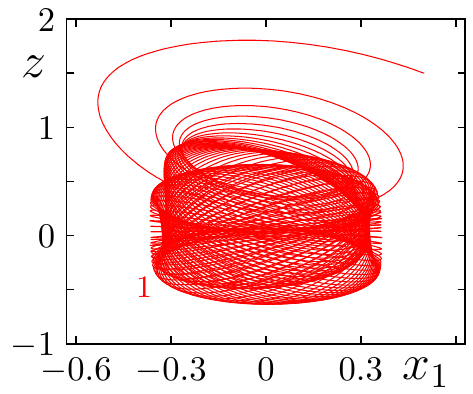} & \includegraphics[width=0.49\columnwidth]{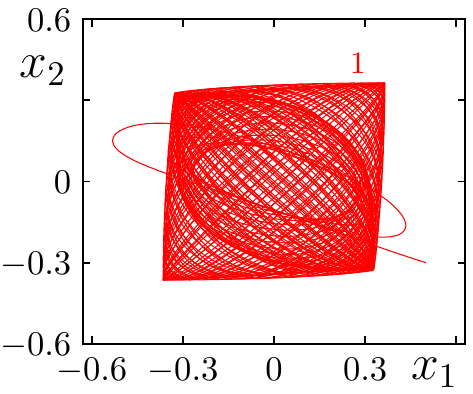} \\
\includegraphics[width=0.49\columnwidth]{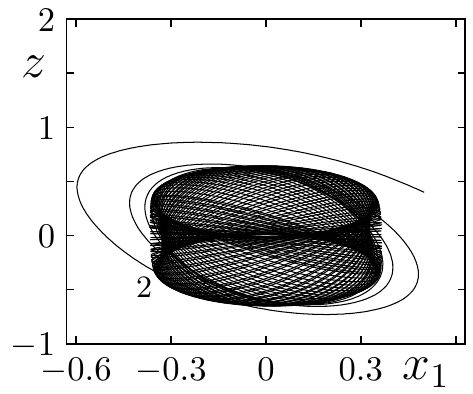} & \includegraphics[width=0.49\columnwidth]{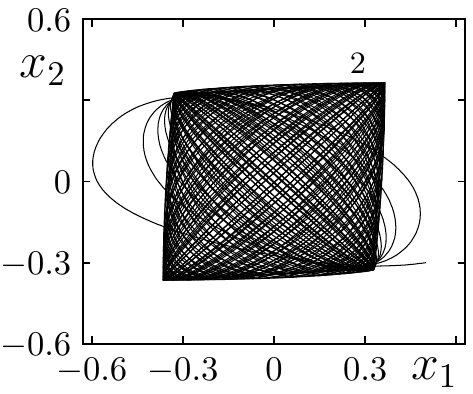} \\
(a) & (b) \\
\end{tabular}
\end{center}
\caption{Phase trajectories of the system (\ref{system}) with the modified memristor state equation (\ref{equation-z2}) in the ($x_{1}$,$z$) (the panel (a)) and ($x_{1}$,$x_{2}$) (the panel (b)) planes. Initial conditions are $x_{1}(0)=0.5$, $y_{1}(0)=0.5$, $x_{2}(0)=-0.3$, $y_{2}(0)=-0.1$, $z(0)=1.5$ (curve 1 in upper panels) and $x_{1}(0)=0.5$, $y_{1}(0)=0.5$, $x_{2}(0)=-0.3$, $y_{2}(0)=0.1$, $z(0)=0.4$ (curve 2 in lower panels). Other parameters are $p=1.05$, $k=0.1$,  $\gamma=0.1$, $a=0.02$, $b=0.8$, $\delta=0.01$.}
\label{fig6}                                                                                                   
\end{figure*}

In order to reveal the influence of the  additional term $- \delta z$ in the memristor state equation, the system (\ref{system}) with the state equation (\ref{equation-z2}) has been considered in numerical experiments at $\delta=0.01$. Figure \ref{fig6} illustrates results of numerical modelling on the example of projections of the phase trajectories obtained from different initial conditions and for the fixed parameters $\gamma=0.1$, $p=1.05$, $k=0.1$. Two sets of the initial conditions were used. The first one ($x_{1}(0)=0.5$, $y_{1}(0)=0.5$, $x_{2}(0)=-0.3$, $y_{2}(0)=-0.1$, $z(0)=1.5$) corresponds to the regime of synchronization in the system (\ref{system}) with the last equation $\dot{z}=x_{1}-x_{2}$ (see the red trajectory 1 in Fig. \ref{fig3}), while the second one ($x_{1}(0)=0.5$, $y_{1}(0)=0.5$, $x_{2}(0)=-0.3$, $y_{2}(0)=0.1$, $z(0)=0.4$) induces the quasi-periodic dynamics (see the black trajectory 2 in Fig. \ref{fig3}). In a case of the system (\ref{system}) with the modified last equation (\ref{equation-z2}) both sets of the initial conditions as well as any other set of the initial conditions give the same oscillatory regime tracing a quasi-periodic attractor. However, one can observe a long transient process for an extremely small value of the parameter $\delta$. The character of the transient process and its duration depend on the initial conditions.
\begin{figure*}[t]
\begin{center}
\begin{tabular}{c}
\includegraphics[width=0.53\columnwidth]{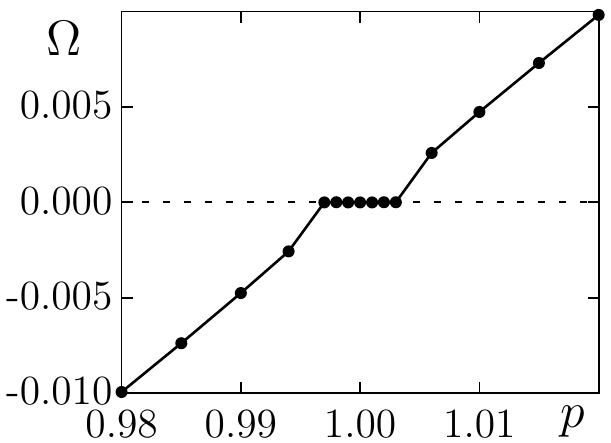}
\end{tabular}
\end{center}
\caption{Dependence of the mean difference frequency $\Omega$ on the frequency mismatch $p$ in the system (\ref{system}) with the modified last equation (\ref{equation-z2}). Parameters are: $\gamma=0.1$, $k=0.02$, $a=0.02$, $b=0.8$, $\delta=0.01$.}
\label{fig7}                                                                                                   
\end{figure*}
The obtained results indicate that addition of the term '$-\delta z$' into the memristor state equation eliminates a continuous dependence of the oscillatory dynamics on the initial conditions.
This fact is proved also by the dependence of the mean difference frequency $\Omega$ on the frequency mismatch $p$ in a case the system (\ref{system}) with the modified last equation (\ref{equation-z2}) [Fig. \ref{fig7}]. All sets of the initial conditions give rise to the same dependence $\Omega(p)$, which means that boundaries of the synchronization area are changeless. 

\section{Conclusions} 
Studying the model of two Van der Pol self-oscillators interacting through memristive coupling has shown intrinsic peculiarities of phase-frequency synchronization. The distinctive character of the synchronization is caused by the features of the memristive coupling and is associated with the existence of a line of equilibria in the phase space. In the case of absolutely identical interacting self-oscillators (there is no frequency mismatch) a steady regime corresponds to the in-phase oscillations in the partial systems. Characteristics of the oscillatory regimes depend on initial conditions in the presence of frequency mismatch. Starting from different initial conditions one can realize either the phase-frequency locking regime or the quasi-periodical dynamics at the same values of parameters. At the same time the boundaries of the synchronizaton region continuously depend on the initial conditions. The analytical results obtained by means of quasi-harmonic reduction have confirmed the numerical data. Consequently, the presence of memristive coupling leads to a special kind of the dynamics and allows to control the effect of synchronization by changing the initial conditions. It has been shown that the addition of a small term into the memristor state equation results in the disappearance of the line of equilibria and destroys dependence of the synchronization on the initial conditions. 

\nonumsection{Acknowledgements}
We are grateful to Galina Strelkova for helpful discussions.
This work was supported by DFG in the framework of SFB 910 and by the Russian Ministry of Education and Science (project code 3.8616.2017/8.9).

%\bibliographystyle{ws-ijbc}
%\bibliography{biblio}

\end{multicols}
\end{document}